# Phantom scalar fields result in inflation rather than Big Rip

A.V. Yurov

Theoretical Physics Department,
Kaliningrad State University, Russia, email yurov@freemail.ru

## Abstract

There exists a variety of exact solutions of the scalar field Einstein equations, allowing for "phantom regions" with the negative kinetic field term. These regions can be cut out, their boundaries being sewn together in such a way that neither the scale factor (along with its first two derivatives) nor density or pressure will experience a jump. Such a domain surgery eliminates the "Big Rip" scenario, substituting for it the standard inflation.

## 1 Introduction

The paper [1] suggested a new concept of death of the universe, called the "Big Rip". Its main idea is to take into account the new hypothetical dark energy components with the state equation $p = w\rho$, with $w < -1$. The matter with this equation-of-state parameter is called "phantom energy" [2]

For the flat FLRW universe this yields the scale factor as a function of cosmic time (in the synchronous reference frame) as follows [1]:

$$\frac{a(t)}{a(t_0)} = \left[\frac{\alpha^2}{2}\sqrt{3\rho(t_0)}\,(t_* - t)\right]^{-2/3\alpha^2},$$

where $\alpha^2 = -1 - w$, $t_0$ is a fixed time instant, and $\rho(t)$ is the density. Clearly as $t \to t_*$ the scale factor blows up, underlying the Big Rip.

The scalar field, minimally connected with gravitation and describing the hypothetical phantom component of dark energy has a negative kinetic term. Quantum theory of such fields is studied in [3], see also the references therein. Analysis has shown that fundamental phantom fields are an impossibility. However effective models of this sort, yet problematic, should not be ruled out. For example, such terms may arise in supergravity [4], in higher-derivative-gravity theories [5]; see also stringy phantom energy [6] and brane-world phantom energy [7].

In this note we attempt to demonstrate that the Big Rip can apparently be ruled out for purely classical (rather than quantum) reasons. There exists a class of exact solutions of the Einstein equations, allowing the "phantom regions". It will be further shown that all these regions can be cut out, with the edges being newly sewn up in such a way that neither the scale factor (and its first two derivatives) nor density or pressure experience a jump.

---

[1] Everywhere in the sequel the common in cosmology system of units with $8\pi G = c = 1$ is used.

## 2 Main idea

Consider the simplest flat universe ($k = 0$) model with the FLRW metric:

$$ds^2 = dt^2 - a(t)^2 \left(dx^2 + dy^2 + dz^2\right),$$

filled with a scalar field with the Lagrangian

$$L = \frac{1}{2} g^{\mu\nu} \partial_\mu \phi \partial_\nu \phi - V(\phi).$$

It's easy to verify that in the case when $\phi = \phi(t)$, the Einstein equations with the cosmological term

$$R_{\mu\nu} - \frac{1}{2} g_{\mu\nu} R = T_{\mu\nu} + g_{\mu\nu} \Lambda,$$

are reduced to the Schrödinger equation

$$\ddot{\psi} = 3 \left(V + \Lambda\right) \psi, \tag{1}$$

with $\psi = a^3$. It is assumed that $V = V(\phi(t)) = V(t)$, while the field time dependence is described by the following equation:

$$\dot{\phi}^2 = -\frac{2}{3} \frac{d^2}{dt^2} \log \psi = -2 \dot{H}. \tag{2}$$

Super-inflation (i.e. the case of dominating phantom fields) comes into play if $\dot{H} > 0$, i.e. when $\ddot{a} > \dot{a}^2 / a$. Suppose the function $\psi(t)$ has a point singularity at $t_*$:

$$\psi(t) \sim (t_* - t)^{-\alpha^2}.$$

Then it follows from (2) that in some neighborhood of $t_*$ the kinetic term becomes negative, i.e. there exists an instant $t = t_1 < t_*$ when $\dot{\phi}(t_1) = 0$. Then $t > t_1$ marks the phantom region, where the kinetic term is negative. To this effect, there are two possibilities: either this region stretches up to infinity into the future, or there exists an instant $t_2$ to the right of the singularity $t_*$, such that

$$\dot{\phi}(t_1) = \dot{\phi}(t_2) = 0, \qquad \dot{\phi}^2(t) < 0, \ t_1 < t_* < t_2.$$

It is easy to convince oneself that the latter alternative takes place if (i) the cosmological constant is zero and (ii) the self-action potential fairly quickly (for instance, as some power, see the next section) vanishes as $t \to \infty$ [2]. Indeed, $H$ Solves the Ricatti equation

$$\dot{H} = V(t) + \Lambda - 3H^2,$$

---
[2]More generally one may have $V(t) + \Lambda \to 0$ fast enough as $t \to \infty$, but such a condition implies adjusting the cosmological constant $\Lambda = -V_{as}$, which hardly appears to be reasonable. Presumably, it is more sensible to render $\Lambda = 0$, the more so as that as of today, there would be no explanation for a small, but nonzero value of $\Lambda$.

that is for $\Lambda = 0$, $V \to 0$ one has $\dot{H} \to -3H^2 < 0$, i.e. the phantom energy domination region is not present over large times. Then the quantity $\dot{\phi}$ must possess another zero, corresponding to some $t_2 > t_*$, q.e.d. [3]

The above conditions being satisfied, the universe would generally speaking enter the inflation phase as $t \to t_1$. Indeed, let us assume that it is born at some time instant $t_0 < t_1$ and suppose there exists a time interval, during which all the energy conditions are satisfied. One can demonstrate that as time goes to $t_1$, it is the strong energy condition to be violated first, the weak one being extant. As a result, the universe would get inflated up to the instant $t_1$. Furthermore, when $t = t_1$ we arrive precisely at De Sitter's phase, but only *instantaneously*. Further follows the "phantom region", **which should be cut out** by matching the solutions corresponding to $t = t_1$ and $t = t_2$. Note that gluing the solutions in such a way should take place under constraints more stringent than those characteristic of similar problems of non-relativistic quantum theory based on the Schrödinger equation. The discussed problem calls for a $C^2$ gluing, namely the functions are to be matched up to their second derivatives. The latter requirement is to eliminate the possibility of discontinuities of pressure and density. A general solution of equation (1) is characterized by two integration constants, and it is not obvious that three matching conditions can be satisfied. Nevertheless, in is easy to verify that the third condition, namely the equality of the second derivatives, is **automatically fulfilled!** provided that the first two, namely the $C^1$ matching has been done. This apparent fluke is due to the fact that the matching is done at an inflection point of $\log \psi$.

The outcome is a solution without jumps of the scale factor and its first two derivatives, the density and its first derivative, and the pressure. Higher order derivatives of $a$ (starting form the third one) $\rho$ (starting from the second) and $p$ (starting from the first derivative) do have jumps however. Pressure and density are connected via an algebraic state equation, and therefore the evolution of $p(t)$ can be described in terms of that of $\rho(t)$. In turn, $\rho(t) = 3\dot{a}(t)^2/a(t)^2$. Thus the whole evolution can be computed from the state equation on the basis of the equation of motion

$$\ddot{a} = -\frac{1}{6}\left(3\left(\frac{\dot{a}}{a}\right)^2 + 3p(a, \dot{a})\right)a.$$

This equation is strongly non-linear, but is nevertheless the second order ODE determined by the initial data. As one can see, the discontinuities pertain to higher order derivatives, which do not enter the above equation. Thus this equation plus the state equation do determine the dynamical evolution of the universe unambiguously, without the region of phantom field domination.

Therefore, the existence of phantom regions entails cosmologies with standard inflation, rather than the universal Doomsday. If all along the solutions possess the right asymptotics (e.g. a Two thirds law dynamics) then the universe would automatically exit the inflationary phase without any extra arrangements. Let us further illustrate it by an easy example.

---

[3]Note that the exit from the phantom region would also take place for negative values of the cosmological constant and a vanishing potential. No fine tuning is required in this case. However (mayhap erroneously!) we prefer to render $\Lambda = 0$.

## 3 Model

The law $a \sim t^{2/3}$ results in a potential

$$V = \frac{2}{3t^2}. \tag{3}$$

The general solution of equation (1) with potential (3) and a zero cosmological constant is

$$\psi^{(-)} = -\frac{c_1}{t} - c_2 t^2, \tag{4}$$

where $c_{1,2} > 0$ are integration constants [4]. The solution describes two different universes, see Fig. 1[5]. The first universe stretches into an unlimited past and has a finite order singularity at a point $t_0 = -(c_1/c_2)^{1/3}$. The second one looks rather mysterious: it emerges from a singularity $t = t_0$, then expands and experiences super-inflation, that is by the time $t = 0$ the scale factor blows up, causing the Big Rip. Further follows a contraction, with a transition to a finite expansion, the scale factor changing the sign. Naturally this solution cannot be taken immediately for what it is. That is the super-inflation phantom region (to the left and to the right of the instant $t = 0$) should not be allowed, as the weak energy condition is violated therein. It is not difficult to verify that the pressure and density have the form

$$p = -\frac{c_1\left(c_1 + 4c_2 t^3\right)}{t^2\left(c_1 + c_2 t^3\right)^2}, \qquad \rho = \frac{\left(2c_2 t^3 - c_1\right)^2}{3t^2\left(c_1 + c_2 t^3\right)^2},$$

and therefore the phantom region is confined to an interval from $t_1$ to $t_2$, where

$$t_1 = \left(\frac{4 - 3\sqrt{2}}{2}\frac{c_1}{c_2}\right)^{1/3} \sim -0.495\left(\frac{c_1}{c_2}\right)^{1/3}, \qquad t_2 = \left(\frac{4 + 3\sqrt{2}}{2}\frac{c_1}{c_2}\right)^{1/3} \sim 1.6\left(\frac{c_1}{c_2}\right)^{1/3}.$$

Clearly, the solution on the right has to be determined by other integration constants:

$$\psi^{(+)} = \frac{b_1}{t} + b_2 t^2.$$

One can show that in order to sew up the solution consistently, one should take

$$b_1 = \left(3 - 2\sqrt{2}\right) c_1 > 0, \qquad b_2 = \left(3 + 2\sqrt{2}\right) c_2 > 0.$$

However, it is not entirely satisfactory as $t_1 \neq t_2$. It is not difficult to do a time shift within $\psi^{(\pm)}$, in order to ensure the following two conditions:
1. The initial singularity corresponds to $t = 0$.
2. $t_1 = t_2$, i.e. there is no time discontinuity.
  Skipping the calculation, here is the answer:
1) For $0 < t \leq t_1$ the dynamics is described as follows:

$$\psi^{(-)}(t) = -\frac{c_1^3}{t - c_1/c_2} - c_2^3 \left(t - c_1/c_2\right)^2, \qquad V^{(-)} = \frac{2/3}{\left(t - c_1/c_2\right)^2}.$$

---

[4] The sign of $\psi$ (and scale factor) is inessential: if $a < 0$ then one must multiply it by $-1$. This because $\psi(t)$ is a solution of linear equation. If $t = t_0$ is the zero of function $\psi(t)$ and $\psi(t < t_0) < 0$ and $\psi(t > t_0) > 0$ (or vice versa) then we have singularity at $t = t_0$ so $\psi(t < 0)$ and $\psi(t > 0)$ describes two different universes.

[5] All the figures are relegated to the Appendix.

2) For $t_1 < t$ it is

$$\psi^{(+)}(t) = \frac{c_1^3(3+2\sqrt{2})}{t-g} + c_2^3(3-2\sqrt{2})(t-g)^2, \qquad V^{(+)} = \frac{2/3}{(t-g)^2},$$

where

$$t_1 = \frac{c_1}{c_2}\left(1 + \sqrt[3]{2 + \frac{3}{\sqrt{2}}}\right) \sim 2.6\frac{c_1}{c_2}, \qquad g = \frac{c_1}{c_2}\left(1 + \sqrt[3]{16 + 12\sqrt{2}}\right) \sim 4.2\frac{c_1}{c_2}.$$

For convenience, a pair of new constants $c_{1,2}^3$ has been introduced; also $t_2 = t_1$. Matching has already been done:

$$\psi^{(-)}(t_1) = \psi^{(+)}(t_1),\ \dot\psi^{(-)}(t_1) = \dot\psi^{(+)}(t_1),\ \ddot\psi^{(-)}(t_1) = \ddot\psi^{(+)}(t_1),\ V^{(-)}(t_1) = V^{(+)}(t_1),\ \rho^{(-)}(t_1) = \rho^{(+)}(t_1),\ \dot\rho^{(-)}(t_1) = \dot\rho^{(+)}(t_1) = 0,\ p^{(-)}(t_1) = p^{(+)}(t_1).$$

The graph $a(t)$ for the second universe is given in Fig. 2. Note that for $t < t_1$ the curve is described by the formula $a^{(-)} = (\psi^{(-)}(t))^{1/3}$, and for $t > t_1$: $a^{(+)} = (\psi^{(+)}(t))^{1/3}$ ($c_1 = c_2 = 1$). Fig. 3 shows the time dependence of the acceleration $\ddot a$. The kink is located at the instant $t = t_1$.

As one can see, inflation over a finite interval takes place (in the region jutting over the $t$ axis): if $t < t_1$ then

$$\ddot a(t \to t_1) \to 0.127\left(\frac{c_2^7}{c_1^4}\right)^{1/3} > 0.$$

Note that as $t \to \infty$ the potential behaves as

$$V \sim \frac{2}{3}e^{-\sqrt{3}\phi}, \qquad \phi \sim \frac{2}{\sqrt{3}}\log t.$$

Therefore, having removed the phantom region, we have arrived at a simple inflation model with an exit and asymptotic transition to Friedmann's universe expansion regime in the phase of the domination of matter. Clearly, the example above does not describe our Universe, as it lacks the oscillation phase. Nor it has the secondary acceleration. Nevertheless, it clearly indicates that one can use the suggested approach in order to produce in principle model cosmologies with exit, avoiding the parameter adjustment. Nor the hybrid models appear to be required.

Let us recall that the solution describes one more, contracting universe, without a beginning but with an end to it. Solutions of this type look rather exotic. However, one inevitably runs into them, as the equations of General relativity allow changing the direction of time.

## 4 Conclusion

It has been shown above that the appearance of phantom regions does not entail super-inflation and the Big Rip (at least o the classical level). On the other hand, the existence of these regions is quite useful for constructing inflationary cosmologies with exit, without parameter adjustments.

It does not follow that super-inflation is absolutely impossible. However, a naive approach to phantom regions does not appear to suggest the correct scenario for evolution of the Universe. As has been shown, a minor "surgery" eliminates the phantom regions, at least in the context of self-action potentials vanishing at infinity, in the case of a non-positive cosmological constant.

*Acknowledgements.*

The author are grateful to Misha Rudnev for useful comments and for the interest in this work. This work was partially supported by the Grant of Education Department of the Russian Federation, No. E00-3.1-383.

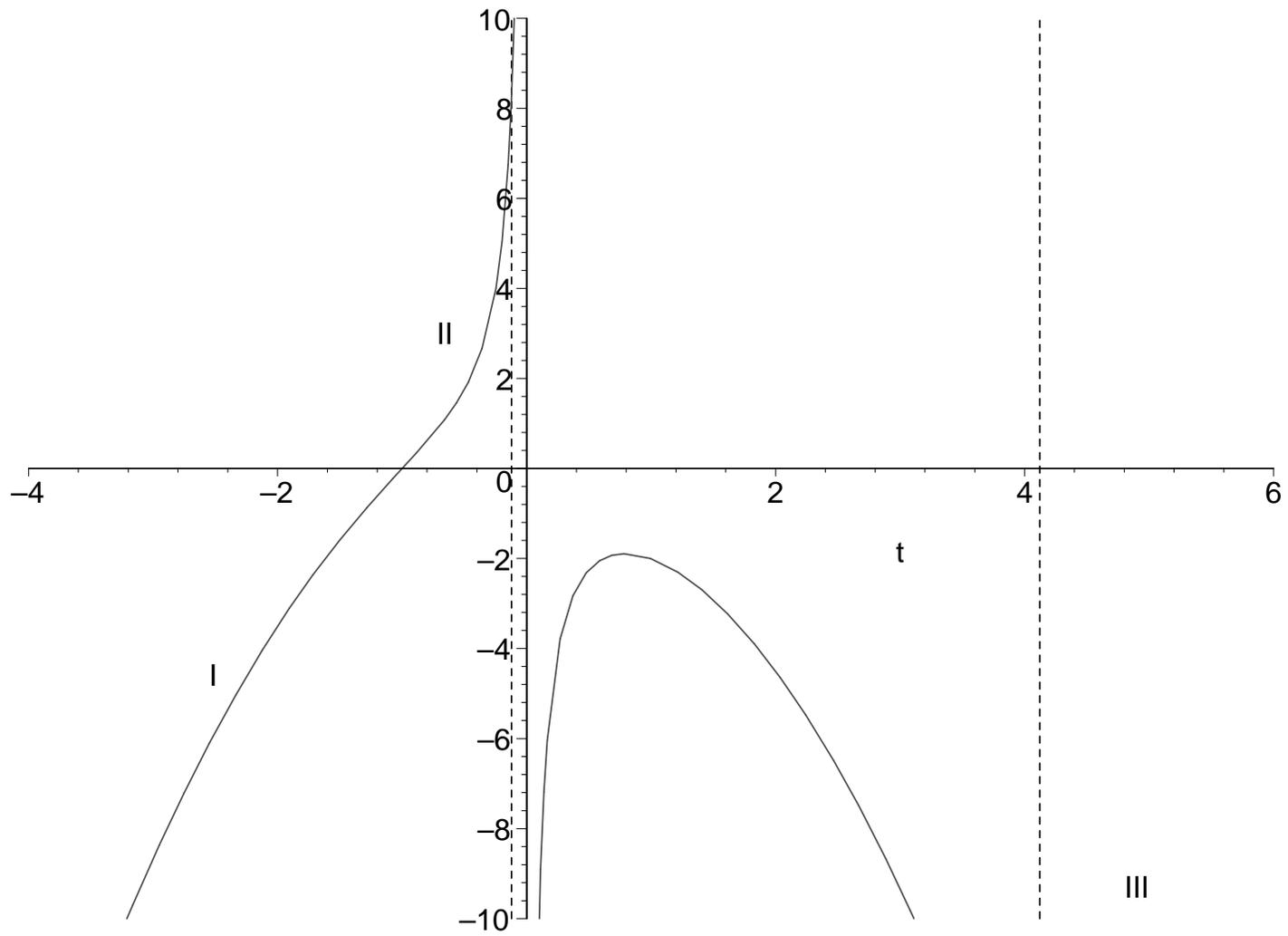

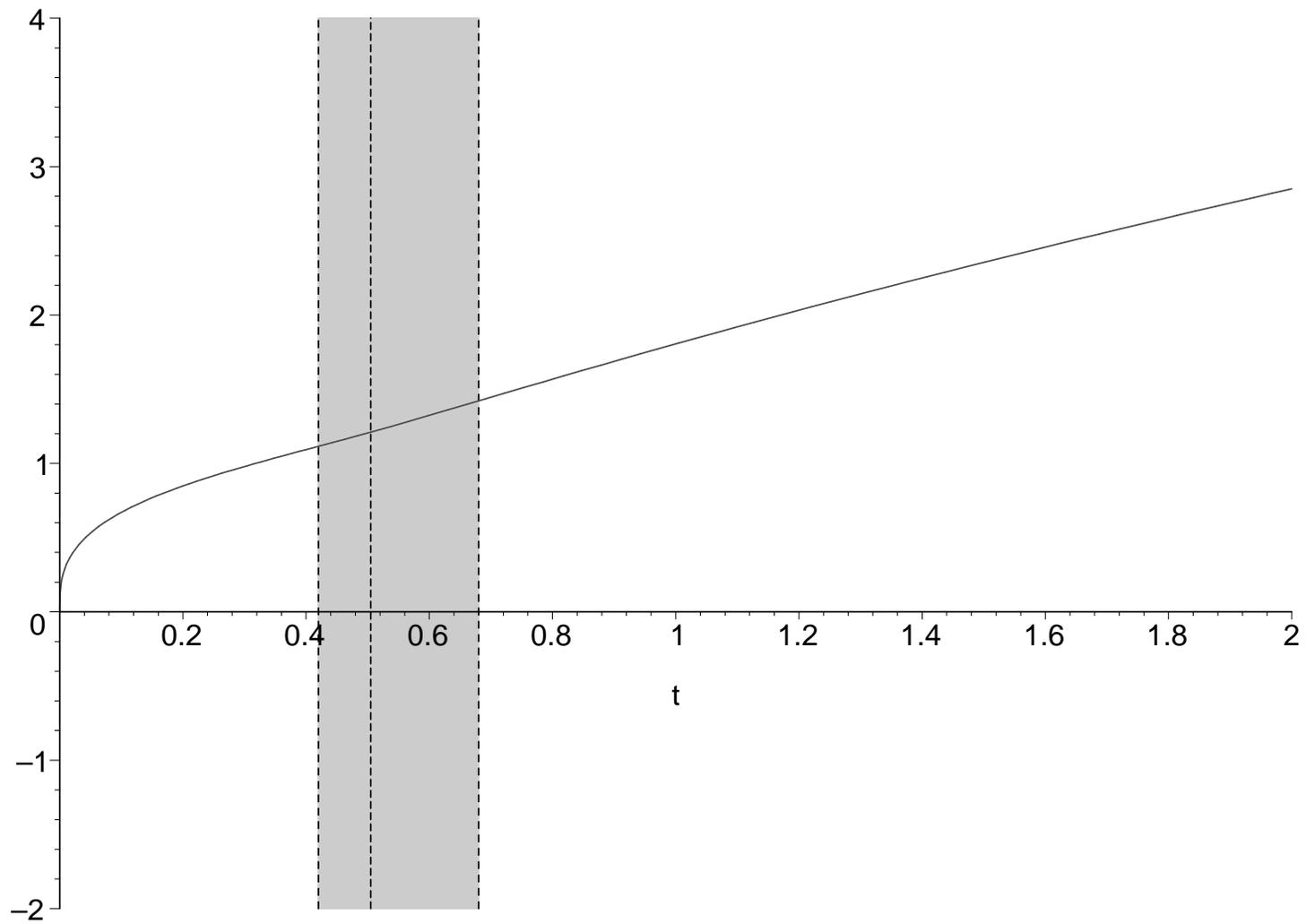

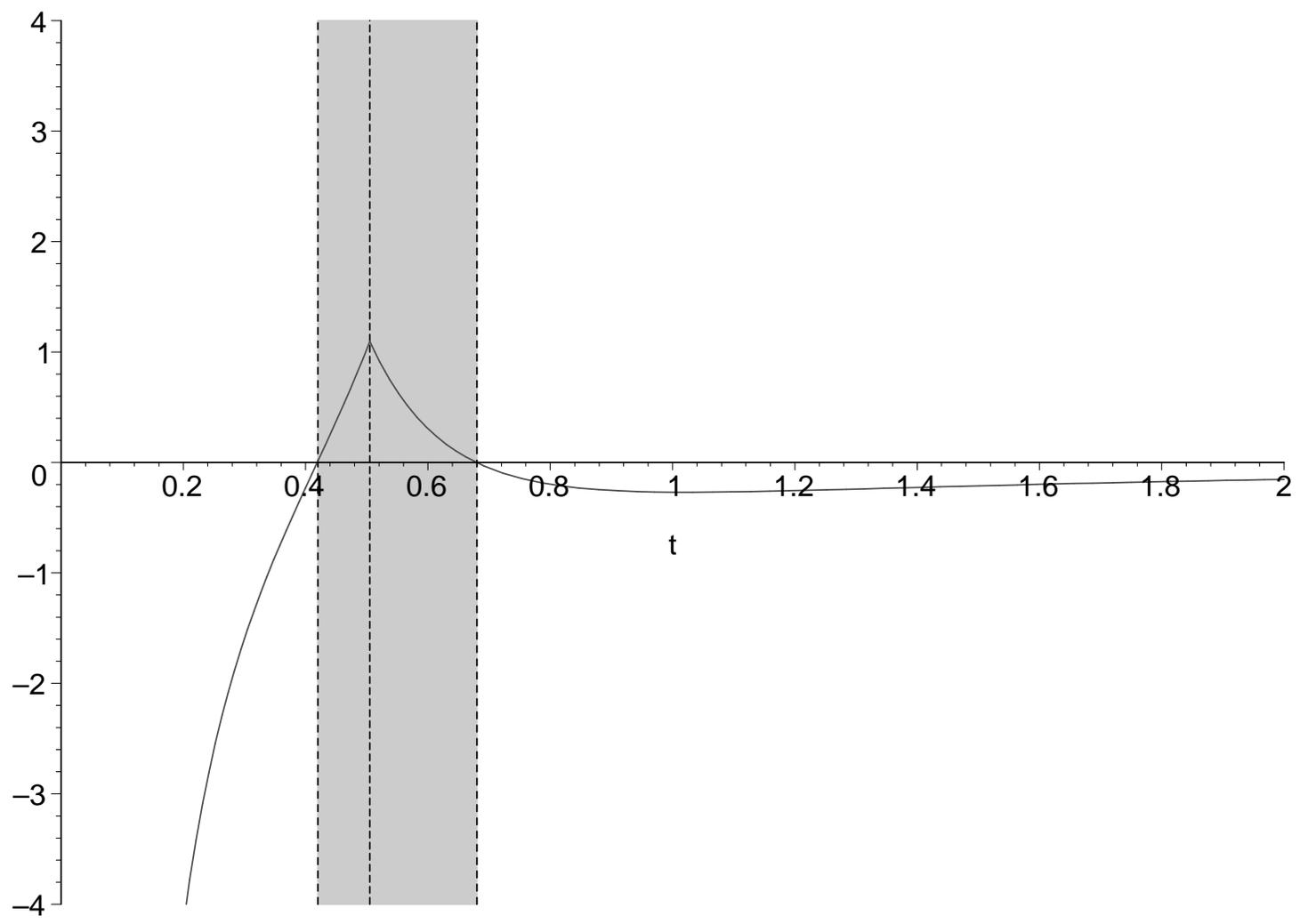